\RequirePackage{fix-cm}
\documentclass[twocolumn,epjc3]{svjour3}  
\smartqed  
\RequirePackage{graphicx}
\RequirePackage[colorlinks,citecolor=blue,urlcolor=blue,linkcolor=blue]{hyperref}
\RequirePackage{cite}
\RequirePackage{booktabs}
\RequirePackage{amsmath}
\RequirePackage{amssymb}

\journalname{Eur. Phys. J. C}
\begin{document}

\title{Machine learning-assisted techniques for Compton-background discrimination in Broad Energy Germanium (BEGe) detector}



\author{G.~Baccolo\thanksref{addr1} \and
        A.~Barresi\thanksref{addr2,addr3} \and
        D.~Chiesa\thanksref{addr2,addr3} \and
        A.~Giachero\thanksref{addr2,addr3} \and
        D.~Labranca\thanksref{addr2,addr3} \and
        R.~Moretti\thanksref{addr2,addr3,e1} \and
        M.~Nastasi\thanksref{addr2,addr3} \and
        A.~Paonessa\thanksref{addr2} \and 
        M.~Picione\thanksref{addr2,e2} \and 
        E.~Previtali\thanksref{addr2,addr3} \and
        M.~Sisti\thanksref{addr3}
}

\thankstext{e1}{e-mail: roberto.moretti@mib.infn.it}
\thankstext{e2}{e-mail: m.picione@campus.unimib.it}



\institute{Dipartimento di Scienze, Università degli Studi di Roma Tre, Rome 00146, Italy \label{addr1}
\and Dipartimento di Fisica "G. Occhialini", Universit\`{a} di Milano-Bicocca, Milan 20126, Italy \label{addr2} \and
Istituto Nazionale di Fisica Nucleare (INFN), Sezione di Milano-Bicocca, Milan 20126, Italy \label{addr3}
}

\date{Received: date / Accepted: date}

\maketitle

\begin{abstract}
High Purity Germanium (HPGe) detectors are powerful detectors for gamma-ray spectroscopy.
The sensitivity to low-intensity gamma-ray peaks is often hindered by the presence of Compton continuum distributions, originated by gamma-rays emitted at higher energies. This study explores novel, pulse shape-based, machine learning-assisted techniques to enhance Compton background discrimination in Broad Energy Germanium (BEGe\texttrademark{}) detectors. We introduce two machine learning models: an autoencoder-MLP (Multilayer Perceptron) and a Gaussian Mixture Model (GMM). These models differentiate single-site events (SSEs) from multi-site events (MSEs) and train on signal waveforms produced in the detector. The GMM method differs from previous machine learning efforts in that it is fully unsupervised, hence not requiring specific data labelling during the training phase. Being both label-free and simulation-agnostic makes the unsupervised approach particularly advantageous for tasks where realistic, high-fidelity labeling is challenging or where biases introduced by simulated data must be avoided.

In our analysis, the full-energy Peak-to-Compton ratio of the ${}^{137}$Cs, a radionuclide contained in a cryoconite sample, exhibits an improvement from 0.238 in the original spectrum to 0.547 after the ACM data filtering and 0.414 after the GMM data filtering, demonstrating the effectiveness of these methods. The results also showcase an enhancement in the signal-to-background ratio across many regions of interest, enabling the detection of lower concentrations of radionuclides.

\end{abstract}
%
\noindent
%

\section{Introduction}\label{sec:intro}
Gamma-ray spectroscopy with High Purity Germanium (HPGe) detectors finds many applications in a wide range of physics fields. HPGe detectors in a low-back\-ground configuration~\cite{Neder2000,Hult2018,Heusser2015} are powerful tools for detecting low concentrations of artificial or natural radionuclides in different types of samples~\cite{Mouchel1992}. Particle physics experiments searching for rare events heavily depend on the choice of radiopure materials~\cite{Azzolini2019, Alduino2017, Haufe2022, Wang2016, Aprile2022}. Germanium detectors are essential to assess the radio-purity of several materials used for such experiments~\cite{vonSivers2015, vonSivers2016}. Improving their sensitivity is thus of primary importance. Environmental samples usually contain trace amounts of various radionuclides, both natural and artificial, which can be key indicators of pollution, environmental changes, or geological pro\-cess\-es~\cite{Woelders2018, Baccolo2023}. Moreover, HPGe are also used in environmental sciences to analyze stable trace elements through neutron activation\cite{shotyk2002, baccolo2017}. According to this technique, the elemental composition of a sample is determined by exposing it to a neutron flux, which induces radioactive activation of specific isotopes depending on their neutron capture cross section~\cite{greenberg2011}. The gamma rays emitted from the activated isotopes are detected and analyzed by HPGe. Enhancing the sensitivity of such methods would lead to improved identification of both stable trace elements and radionuclides, with beneficial effects on a series of disciplines, from biogeochemistry, to paleoclimatology and analytical chemistry.

For setups placed in ground-level laboratories, conventional passive shielding is usually not sufficient to achieve sensitivity below hundreds of mBq/kg. In such cases, it is crucial to develop active shields (such as muon vetoes) ~\cite{Baccolo2021} and Compton suppression systems ~\cite{Semkow2002,deOrduna}, to enhance sensitivity and effectively mitigate the influence of external sources of radiation. These techniques enable the improvement of overall sensitivity, allowing for the detection of lower concentrations of artificial or natural radionuclides in materials. Specifically, in gamma spectra, the presence of gamma-ray lines emitted by radionuclides in relatively small concentrations could be hidden by the Compton continuum. The incomplete deposition of gamma-ray energy in a detector increases the Compton background counts in spectra and reduces the peak-to-Compton ratio, adversely affecting qualitative and quantitative spectral analyses. Anticoincidence techniques are commonly employed for Compton discrimination. Nevertheless, these approaches require one or more anticoincidence detectors surrounding the HPGe detectors. This leads to increased system volume, electronic and equipment complexity, and maintenance costs.

A different approach for reducing the Compton background in HPGe detectors is to apply pulse shape discrimination (PSD) algorithms to the acquired data. A single-site event (SSE) originates from a single interaction of a gamma ray within the crystal, resulting in a localized energy deposition in a small region of less than 1\,mm$^3$. In contrast, a multi-site event (MSE) is generated by multiple interactions occurring at several locations within the crystal, resulting in multiple energy depositions separated by some millimetres or centimetres. In HPGe gamma-ray spectra, Compton events, photoelectric absorption with X-ray escape, and pair production with double escape are mainly SSEs, whereas full-energy peaks are composed mostly of MSEs. The discrimination and rejection of SSE occurrences thus allow the reduction of the continuum background (when dominated by the Compton continuum) while preserving most of the peak areas, so that it becomes possible to enhance the signal-to-background ratio. This, in turn, improves the overall sensitivity and accuracy of the measurements. Since the 1960s,  PSD analysis in HPGe detectors has shown the possibility to discriminate between energy deposition in SSEs and MSEs~\cite{deOrduna}. HPGe detectors with a semi-planar crystal geometry and a small readout electrode, also known as p-type point contact (PPC) detectors, have proven to be very suitable for pulse shape discrimination~\cite{diVacri2009,Budjas2009}. This configuration not only provides excellent energy resolution, but also an efficient charge collection process that results in different pulse shapes between SSEs and MSEs, thus allowing enhanced PSD capabilities with respect to coaxial HPGe detectors.

The goal of the presented work is to investigate how pulse shape discrimination based on a machine learning (ML) approach could increase sensitivity in gamma-ray spectroscopy, meeting the requirements of the aforementioned applications. This approach is developed using measurements from calibration sources and environmental samples.
To demonstrate the potential of the machine-learning-based analysis presented here, we used a Broad Energy Germanium (BEGe\texttrademark{}) detector, a commercial type of PPC detector.

Efforts have been made to achieve MSE and SSE classification through ML models \cite{HollAutoencoderBEGe, MisiaszekMLP}, leveraging Multilayer Perceptrons (MLP) \cite{HORNIK1989359} and Autoencoders \cite{Schmidhuber_2015} for deep pulse-shape elaboration. Although promising, the performance of such supervised learning approaches heavily depends on the quality and representativeness of an already existing labeled training dataset. Obtaining such a dataset involves either using simulated waveforms, which may introduce systematic differences compared to real data, or relying on \textit{a priori} information. For instance, assigning labels to specific intervals in the gamma energy spectrum can lead to misclassification due to partial overlap between classes (e.g., SSE and MSE events that deposit a similar amount of energy in the detector).

In this study, we propose two distinct classification approaches. The first one involves an Autoencoder-MLP architecture, that we refer to as the Autoencoder Classifier Model (ACM). The second approach utilizes the Gaussian Mixture Model (GMM) \cite{GMM}, an unsupervised clustering method that can be applied directly to unlabeled datasets with minimal feature processing. In this work, we adapt the ACM to our specific detector and calibration dataset, with the primary purpose of providing a benchmark for evaluating the performance of the GMM, serving as a comparison against an already demonstrated \cite{HollAutoencoderBEGe}, well-functioning methodology.

In Sec. \ref{sec:setup} and \ref{sec:exp}, we provide a description of the detector in use and the data collection process. Sec. \ref{sec:acm}  and \ref{sec:gmm}  detail the machine learning approaches and the necessary individual preprocessing and calibration steps. We compare the performance of the two models on real data in Sec. \ref{sec:cryoconite}, demonstrating their SSE rejection capability and ability to retrieve peak information from an unknown radioactive sample. Finally, our conclusions and a discussion of possible future directions are presented in Sec. \ref{sec:conclusions}.

\section{Experimental setup}\label{sec:setup}

\begin{figure}[!t] 
\centering\includegraphics[width=0.45\textwidth,clip]{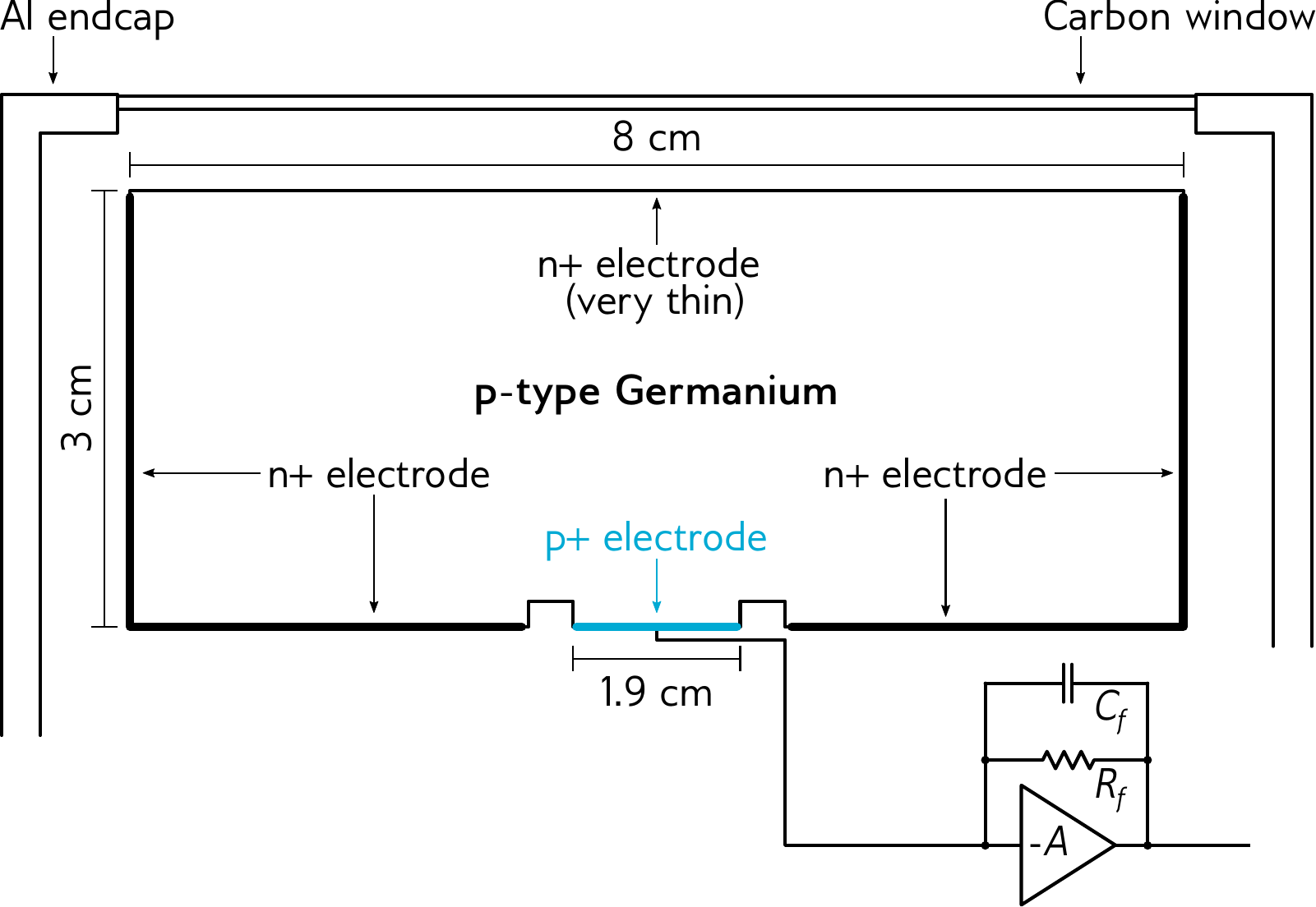}
\caption{Vertical cross-section of the BEGe\texttrademark{} detector crystal model BE5030, with an approximate representation of the electrodes. The positive external electric potential is applied to the n\textsuperscript{+} electrode, while the signal is taken from the p\textsuperscript{+} electrode. The diagram of a charge-sensitive preamplifier is also shown.}
\label{fig:detector}       
\end{figure}

The HPGe detector used in this work is a commercial BEGe\texttrademark{} detector (model BE5030~\cite{MirionBEGe}) manufactured by CANBERRA, now part of Mirion Technologies, Inc. A schematic view of the detector configuration is shown in Fig. \ref{fig:detector}. The p-type germanium crystal has a semi-planar geometry, with a diameter of 8\,cm and a height of 3\,cm. The signal contact is provided by a small boron-implanted p\textsuperscript{+} electrode with a diameter of 1.9\,cm, while the high-voltage contact is provided by a lithium-diffused n\textsuperscript{+} electrode, which extends over most of the crystal's remaining surface.
The crystal is mounted in a copper cup placed in an
aluminium endcap, whose thin entrance window made of composite carbon allows the detection of low-energy gamma rays.

\begin{figure}[!t] 
\centering\includegraphics[width=0.4\textwidth,clip]{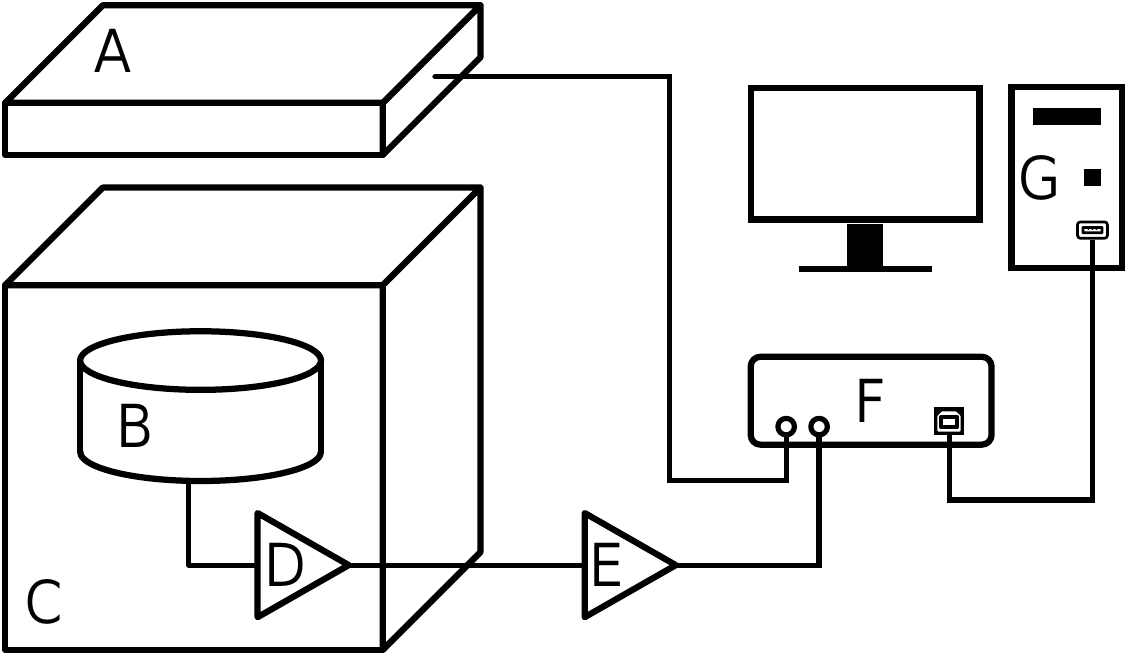}
\caption{Block diagram of the BEGe\texttrademark{} detector read-out system: (A) Plastic scintillator, (B) BEGe\texttrademark{} detector, (C) Passive shielding, (D) Charge-sensitive preamplifier, (E) Amplifier, (F) CAEN DT5725S digitizer, and (G) control computer with the acquisition software.}
\label{fig:readout}       
\end{figure}

A block diagram of the complete system, which is located in a shallow underground laboratory at the University of Milano-Bicocca, is illustrated in Fig.~\ref{fig:readout}. The detector is placed in a 15\,cm-thick shielding composed of rectangular lead bricks
arranged in multiple layers to minimize the probability of gamma rays from the environment passing through potential cracks. The inner side, which faces the detector, is covered with a 5\,cm-thick layer of copper to absorb photons generated by physics processes (such as bremsstrahlung) occurring within the shielding itself.
The detector is also equipped with a standard charge-sensitive preamplifier, which produces a linear tail pulse for each gamma ray event happening in the crystal, whose amplitude is proportional to the energy deposited by the gamma-ray in such event. After reaching the maximum amplitude, the pulse decays exponentially with a time constant of tens of microseconds. Moreover, the leading edge of the pulse reflects the time characteristics of the charge collection process within the crystal, which happens in a timescale of some hundreds of nanoseconds; in particular, its shape mirrors the time profile of the charge induced on the p\textsuperscript{+} signal contact following one or more gamma-ray interactions.



The output of the BEGe\texttrademark{} preamplifier is sent to a custom low-noise non-shaping amplifier, which provides a voltage amplification stage
without altering the signal shape. This allows for optimal utilization of the input dynamics of the subsequent digitizer
while preserving the information contained in the pulse, which is necessary for
the machine learning analyses proposed in this paper.

A panel-type plastic scintillator with dimensions of 70\,cm x 28\,cm x 5\,cm from the NUVIA NuDET product line~\cite{NUVIAplastic}, placed above the lead shielding, serves as an active shield against cosmic radiation, producing a voltage signal when a particle passes through its active volume. Operating the plastic scintillator and the BEGe\texttrademark{} detector in anticoincidence mode allows the reduction of the background generated by muon-induced events.

Both the non-shaping amplifier and the plastic scintillator outputs are connected to two different inputs of a CAEN DT5725S digitizer~\cite{DT5725S}, whose ADC continuously converts the instantaneous voltage value at each input into a proportional
digital number.
The digitizer then performs two important functions on each stream of numbers coming from the ADC. First, it identifies the presence of pulses due to radiation interactions in the corresponding detector, keeping memory of the arrival time of every pulse. The identification of a pulse is performed by a Trigger and Timing Filter, a digital algorithm whose output is a bipolar signal with the zero-crossing independent of the input pulse amplitude~\cite{DPP-PHA}. The presence of a pulse is detected when this signal exceeds a predetermined threshold, and the zero-crossing time is taken as the arrival time at the digitizer of the input pulse. Then, the digitizer shapes each identified digital pulse into a suitable waveform and determines its height. The shaping of an identified pulse is carried out by the Trapezoidal Filter, also known as moving window deconvolution~\cite{DPP-PHA}. This digital filter transforms the typical linear tail pulse generated by a charge-sensitive preamplifier into a trapezoid whose height is proportional to the amplitude of the input signal.

The digitizer is controlled by the CAEN data acquisition software CoMPASS~\cite{CoMPASS}. During a measurement, the digitizer processes the input signals and sends the information extracted by its digital filters, together with each digitized input waveform, to the software. The latter applies the anticoincidence between the \newline BEGe\texttrademark{} and the scintillator pulses to remove muon-induced events, constructs the energy spectrum of the BEGe\texttrademark{} detector and saves all the data to disk for the offline analyses.

The system operations were optimized to achieve the best performance for the intended application. The implementation of the active shield resulted in a reduction of the muon-induced events, lowering the total background rate from 500\,mHz to 375\,mHz within the energy range 30-2700\,keV. The shaping parameters of the Trapezoidal Filter were set at 2\,$\mu$s for the rise time and 1\,$\mu$s for the flat top width, resulting in an optimal energy resolution of 0.85\,keV at 59.54\,keV, 1.35\,keV at 661.66\,keV and 1.80\,keV at 1332.49\,keV.

\section{Experimental Measurements}\label{sec:exp}

The ML approaches employed in this work required the calibration and validation of the models on a large set of data. A source of $^{232}$Th is very convenient for this purpose, since its natural radioactive chain includes the emission of many gamma rays across the entire energy range of interest for most applications in gamma spectroscopy (up to about $3 \, \mbox{MeV}$). Specifically, it provides distinct features in the energy spectrum\,---\,the full-energy, single and double escape peaks, and the single- and multi-Compton regions\,---\,useful to develop the ML models as explained in the following sections. Thus, a standard source of $^{232}$Th with an activity of $3 \, \mbox{Bq}$ was measured for a period of three weeks, in order to obtain a large dataset. For each pulse, in addition to the energy information, the digitizer/software system saved to disk the rising part of the input waveform with a $1500\,\mbox{ns}$ window sampled at 250\,MHz (see Fig.\,\ref{fig:Experimental_events}). The temporal width of the window was chosen in order to save only the needed part of the pulse: its leading edge reflects the time profile of the charge collection process, which was required for the ML-based pulse shape discrimination analyses of this work, while the exponential decay is determined solely by the preamplifier feedback circuit, thus not being of interest for the subsequent processing.

The two ML models were then tested and compared on a four-day-long measurement of a cryoconite sample collected at the Morteratsch glacier in July 2018 (Switzerland). The cryoconite is the dark sediment that accumulates on the melting surface of global glaciers, consisting of a mineral and an organic fraction. One of its features is the ability to accumulate fallout environmental radionuclides with unprecedented efficiency, making this natural sediment one of the most radioactive environmental matrices found on the Earth's surface \cite{Beard2024}, with activities of single fallout radionuclides easily exceeding 1000 Bq kg ${}^1$. The high activity and the presence of several radionuclides make cryoconite a suitable sample to test the ability of the proposed methods to reduce the background related to Compton scattering.

\section{Autoencoder Classifier Model}\label{sec:acm}
In this section, we describe the ACM for SSE-MSE classification. This method relies on two feed-forward Deep Neural Networks, which we will refer to as Autoencoder and Classifier. As mentioned in Sec. \ref{sec:intro}, the effectiveness of Deep Learning for this classification problem in BEGe\texttrademark{} detectors has already been addressed \cite{MisiaszekMLP, HollAutoencoderBEGe}. On top of these results, we tailored a supervised Deep Learning approach to data collected by our detector, providing a viable benchmark for the Gaussian Mixture method, introduced in Sec. \ref{sec:unsupervised}.

\subsection{Dataset preprocessing}
\label{sec:ACMpreprocessing}
As already mentioned in Sec. \ref{sec:intro}, we expect the rate of SSE and MSE to vary significantly with the energy deposition in the detector.
In order to train a robust classifier solely based on the event topology, allowing us to mitigate the background for low-activity peaks which are covered by a large amount of SSE, we prevented our model from exploiting integral energy information. We achieved this by normalising the digitized waveform from $0$ to $1$ for each event.
We note that this approach only partially eliminates other energy-correlated variables, such as the signal-to-noise ratio. To further impose a pulse shape-only discrimination, we trained our model with events in the $[300, \, 2700]\,\mbox{keV}$ energy range, which presents negligible voltage noise with respect to signal intensity.

In addition to voltage rescaling, digitized waveforms were down-sampled by a factor of three, reducing the input array size from $366$ to $122$ entries, encoding voltage over time. Each entry in the new array was calculated as the average of the three corresponding consecutive values in the original array. This allowed us to speed up training without affecting the overall model performance, as the signal features distinguishing the two classes happen at a much larger timescale than the detector temporal resolution (see Fig. \ref{fig:Experimental_events} for reference).

\subsection{Model architecture}
\begin{figure}[h]
    \centering
    \includegraphics[width = 0.45\textwidth]{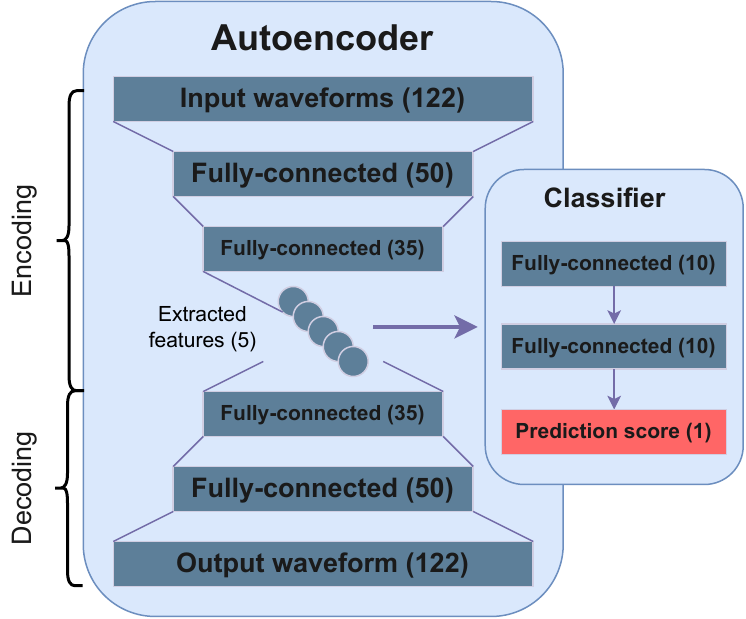}
    \caption{Schematic representation of the Autoencoder Classifier Model. The encoding process of the Autoencoder neural network extracts a compact representation of the preprocessed waveforms, which is then fed as input to the Classifier. The Classifier assigns an MSE/SSE prediction score to each waveform. The Autoencoder and Classifier are trained separately, with the Autoencoder training occurring independently of the Classifier. Neuron numbers are indicated for each layer.}
    \label{fig:acm}
\end{figure}

The ACM model comprises two feed-forward architectures. The first one is an Autoencoder, an unsupervised Neural Network which is able to map input data to a lower-dimensional latent space through an encoding process. After encoding, latent space variables undergo a decoding phase, which reconstructs the original input minimizing the information loss due to compression, as depicted in Fig. \ref{fig:acm}. Both encoding and decoding processes consist of a stack of feed-forward connected layers. The dimension of the latent space corresponds to the number of neurons in the last encoding layer, called the "bottleneck layer".

We trained the model by minimizing a loss function $L_{\text{aut}}$ defined as the mean squared error between the input and output waveforms, corresponding to equation \ref{eq:cost_autoencoder}:
\begin{equation}
    L_{\text{aut}}= \frac{1}{n} \sum_{i=1}^{n}\left(\hat{y}_i - y_i \right)^2
    \label{eq:cost_autoencoder}
\end{equation}
where $n$ is the waveform length, $\hat{y}_i$ is the $i$-th entry of the input voltage and $y_i$ is the $i$-th predicted output voltage.

The second architecture of the ACM model is another feed-forward Neural Network employed as a binary classifier. It operates by accepting the encoded features generated by the Autoencoder model as input and producing a prediction score $\in [0, 1]$. Scores closer to zero are indicative of the SSE class, while values closer to one are associated with MSE. For this classification problem, we chose a standard learning strategy, i.e.\ minimizing the Binary Cross Entropy \cite{zhang2018generalized} cost function described in equation \ref{eq:cost_classifier}. 
\begin{equation}
    L_{\text{class}} = - \frac{1}{m}\sum_{i=1}^{m}  \left( y_i \log{\hat{y}_i} +  (1-y_i) \log{(1-\hat{y}_i)}\right)
    \label{eq:cost_classifier}
\end{equation}
Both Autoencoder and Classifier networks have been trained using the Adam optimizer \cite{kingma2017adam}, and the models' architecture, training and predictions were implemented using the \texttt{TensorFlow} framework \cite{tensorflow2015-whitepaper}, version 2.10.
A scheme of the ACM model is represented in Fig. \ref{fig:acm}.

\subsection{Model calibration with a ${}^{232}$Th dataset}
\label{ACM_calibration}
\begin{figure*}[h]
    \centering
    \includegraphics[width = 0.9\textwidth]{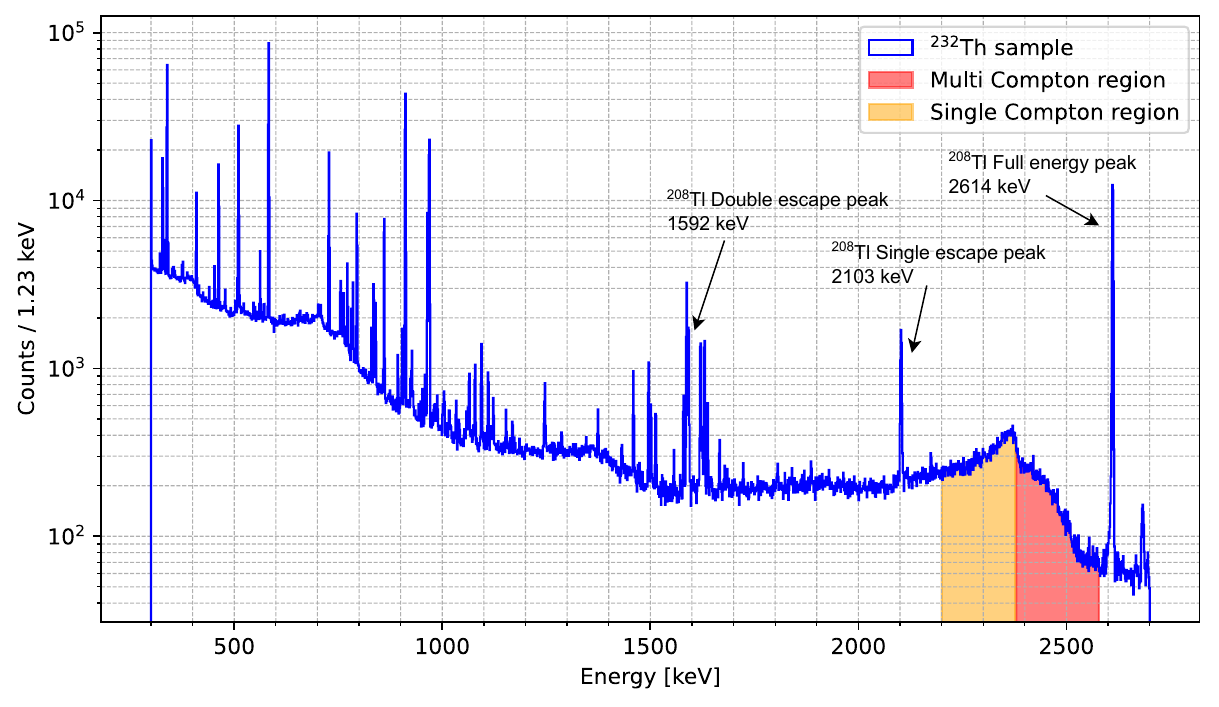}
    \caption{BEGe\texttrademark{} gamma spectrum between $300\,\mbox{keV}$ and $2700\,\mbox{keV}$, in which we underline regions and peaks useful for training a binary Classifier from initially unlabelled data. In the yellow region, we expect to mostly detect photons from the ${}^{208}$Tl single-Compton continuum. In the red region, only multi-Compton events from ${}^{208}$Tl can occur.}
    \label{fig:232Th}
\end{figure*}
For the ACM model training and validation, we used unlabeled events collected by a ${}^{232}$Th radioactive source and processed as described in \ref{sec:ACMpreprocessing}.

The Autoencoder network is agnostic to whether an event belongs to the single- or multi-site class, as it serves as a way to efficiently compress information for the Classifier network. For this reason, we trained the Autoencoder with a subset of uniformly extracted samples from the full dataset. On the other hand, explicit training data labelling was necessary for the Classifier. To accomplish this, we selected events from specific energy regions, where most events were expected to belong exclusively to either the MSE or the SSE class. This distinction was made possible by the abundance of ${}^{208}$Tl, which is a decay product in the ${}^{232}$Th decay series. ${}^{208}$Tl generates a distinct $2614 \, \mbox{keV}$ full-energy peak, along with the associated single- and double-es\-cape peaks, as well as single- and multi-Compton distributions, as shown in Fig. \ref{fig:232Th}. Detailed information regarding the energy intervals used for training the Classifier can be found in Table \ref{tab:training_regions}.

Evaluating the fraction of mislabeled data in each interval would require a major effort in the modeling and simulation of the detector, the radioactive source under consideration and background radiation in our specific environment. The only exception is the interval between $2098$ keV and $2109$ keV, where we expect to collect the most significant fraction of mislabeled events. This is because the dominant event type in the interval is ${}^{208}$Tl single-escape (MSE class), but the subdominant process belongs to the opposite class (SSE), i.e.\ the ${}^{208}$Tl Compton background. By fitting the region with a Gaussian curve plus a constant, we estimated that around $4\%$ of the events belong to the Compton process and, therefore, are mislabeled. This is a clear disadvantage of supervised models for this task that motivates the use of unsupervised methods such as the GMM described in Sec. \ref{sec:gmm}.

\begin{table}
  \centering
    \caption{Energy intervals used for training the Classifier neural network, grouped into MSE and SSE labels. For each energy interval, we report the dominant process we expect to observe.}
  \begin{tabular}{ccc}
    \toprule
    Label & Energy region [keV] & Dominant process \\
    \midrule
    MSE  & $[2609, \, 2621]$  & ${}^{208}$Tl full-peak \\
      & $[2098, \, 2109]$  & ${}^{208}$Tl single-escape \\
      & $[2436,\,2564]$  & ${}^{208}$Tl multi-Compton \\ \hline
    SSE & $[1590,\,1595]$ &  ${}^{208}$Tl double-escape\\
     & $[2130,\,2370]$ &  ${}^{208}$Tl single-Compton\\
    \bottomrule
  \end{tabular}
  \label{tab:training_regions}
\end{table}

We manually tuned the training energy intervals in order to enhance the Classifier accuracy, i.e.\ the fraction of correctly classified events over the total validation sample, when setting a threshold score of $0.5$. In order to avoid class imbalance \cite{balance}, we further subsampled the Classifier training dataset in order to have the same number of events for each class, while keeping the same proportions between energy intervals of the same class as they naturally occurred in the detector.

Using an Autoencoder trained on a larger dataset spanning the full energy range of interest allows for the extraction of a few highly informative features per event. This approach simplifies the Classifier architecture compared to using raw waveforms, reducing the overall degrees of freedom, mitigating overfitting, and enhancing the generalizability of predictions to energy intervals beyond the training regions. Similar benefits were also highlighted in \cite{HollAutoencoderBEGe}.

\begin{figure}
    \centering
    \includegraphics[width = 0.45\textwidth]{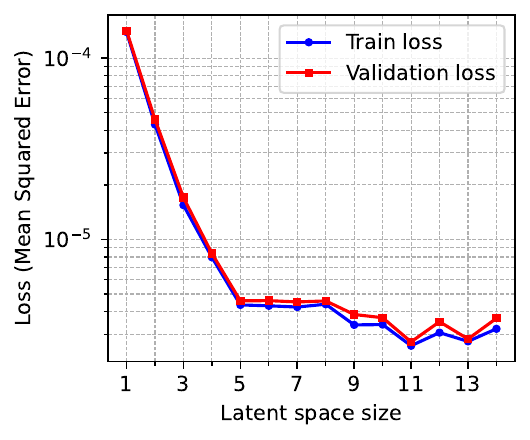}
    \caption{Autoencoder loss function evaluated for the training and validation set at different latent space dimensions, i.e.\ number of neurons in the bottleneck layer. Higher loss values indicate a less accurate reconstruction of the original waveform by the Autoencoder.}
    \label{fig:bottleneck_scan}
\end{figure}

\begin{figure}
    \centering
    \includegraphics[width = 0.49\textwidth]{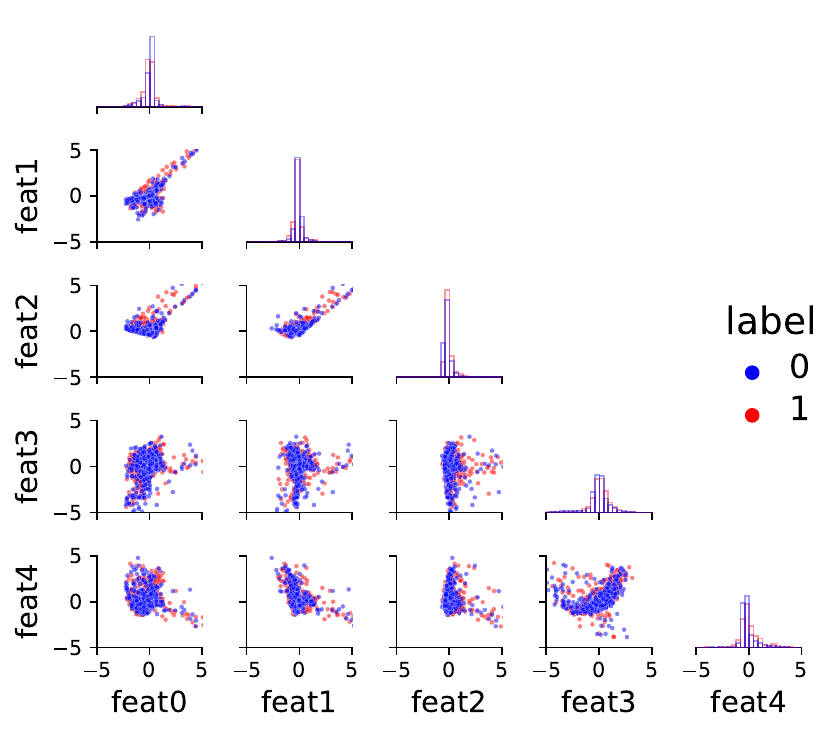}
    \caption{Standardized distributions of features extracted from the Autoencoder model for events in the Classifier training set.}
    \label{fig:feature_distribution}
\end{figure}

The initial phase in the optimization of the ACM architecture entailed establishing the structure of the Autoencoder layer, specifically its depth and size. The evaluation involved an analysis of the Mean Squared Error loss function across different bottleneck sizes to determine an optimal balance between the latent space dimension and the accuracy of waveform reconstruction, without altering the remaining layer architectures. To this end, we selected a configuration comprising two hidden layers before the bottleneck, with sizes (i.e.\ neuron number) of $50$ and $35$ respectively, followed by two hidden layers after the bottleneck, with sizes of $35$ and $50$ respectively, for the decoding process. We employed the LeakyReLU \cite{xu2015empirical} activation functions for every neuron and applied batch normalization to every fully connected step.
The outcome is depicted in Fig. \ref{fig:bottleneck_scan}, using $2\times 10^5$ events for training and $2\times 10^5$ for validation  (the sets are disjoint). Convergence in the Autoencoder training was reached for all bottleneck size configurations after roughly $400$ epochs\footnote{The Autoencoder was trained at a rate of $3.50$ seconds per epoch using Intel\textsuperscript{\tiny\textregistered}Core\texttrademark{} i5-10400 CPU (2.90 GHz).}. Varying the latent space size, we observe an initial rapid decrease in the loss function that plateaus for bottleneck layers with more than $5$ neurons. This trend was robust to variations in the network hyperparameters such as the number of neurons in intermediate layers with no significant differences in the loss function values after reaching convergence. For this reason, we fixed the bottleneck layer size at $5$, ensuring negligible information loss (a Mean Squared Error around $4\times 10^{-6}$ on average per event) by extracting only $5$ features from the input waveforms.
This value was further justified \textit{a posteriori} by training the Classifier network and monitoring the classification accuracy for different bottleneck sizes. For a number of extracted features $>5$, accuracy variations are negligible in comparison with epoch-by-epoch fluctuations in the convergent regime ($\lesssim1\%$ after $200$epochs\footnote{The Classifier was trained at a rate of $0.15$ seconds per epoch using Intel\textsuperscript{\tiny\textregistered}Core\texttrademark{} i5-10400 CPU (2.90 GHz).}).
Fig. \ref{fig:feature_distribution} shows the encoded $5$ feature distributions for the MSE/SSE classes in the labelled subset.
We standardized the $5$ features in input to the Classifier model in order to have zero mean and unit standard deviation. Thanks to the small feature size, the Classifier architecture was elementary, comprising two hidden layers with a LeakyReLU activation function and a single-neuron output layer with a softmax \cite{softmax} activation function. 

\begin{figure}
    \centering
    \includegraphics[width = 0.5\textwidth]{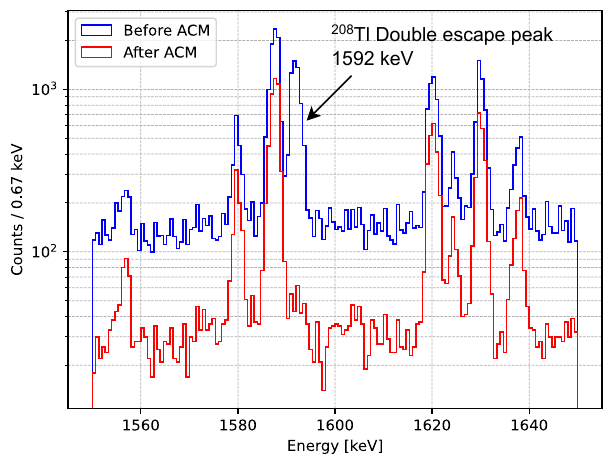}
    \caption{Gamma-ray spectrum between $[1500\,, 1650]\,$keV, before and after employing the ACM model selection with a threshold score set to $0.58$. Notably, the double-escape peak associated with ${}^{208}$Tl is observed to vanish, accompanied by an improvement in the signal-to-background ratio for the remaining peaks between a minimum factor of $1.5$ to a maximum of $2.1$.}
    \label{fig:before_after_ACM}
\end{figure}
\begin{figure*}
    \centering
    \includegraphics[width = 0.95\textwidth]{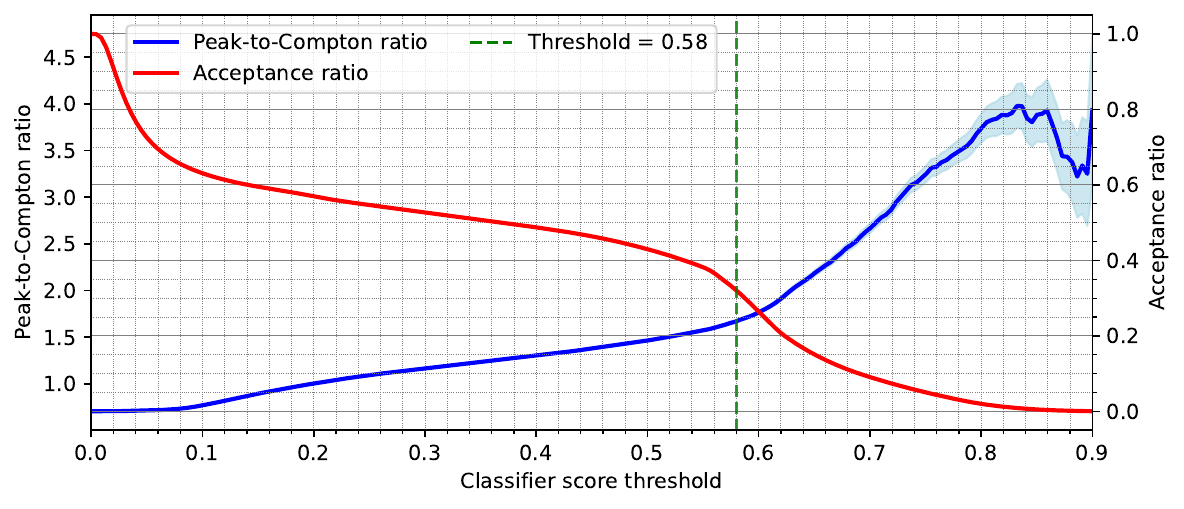}
    \caption{${}^{208}$Tl peak-to-Compton ratio (blue) and the fraction of retained events (red) as a function of the ACM working point. The green dashed line indicates our reference working point of $0.58$, for which we observe an improvement of the peak-to-Compton ratio up to $1.75$ by rejecting around $65\%$ of the total events in the dataset. The light-blue error band accounts for the statistical fluctuation of the peak-to-Compton ratio by propagating the Poissonian uncertainties of counts in the peak and Compton regions.}
    \label{fig:threshold_set}
\end{figure*}

Due to the considerable overlap in the distribution of the two classes, as shown in Fig. \ref{fig:feature_distribution}, the classification performance of the neural network only reaches a modest level, achieving a validation accuracy of approximately $68\%$-$69\%$. Nevertheless, the ACM filtering can significantly enhance the signal-to-background ratios of full-energy and single-escape peaks, while effectively suppressing the double-escape peak associated with ${}^{208}$Tl, demonstrating an effective SSE mitigation \cite{MisiaszekMLP, HollAutoencoderBEGe, gerdaph2}. Fig. \ref{fig:before_after_ACM} depicts the effect of the ACM model on a validation set. This notable outcome is a result of employing a slightly more stringent working point for the classifier, specifically a minimum acceptance score of $0.58$. By adopting this working point, we estimate that approximately $35\%$ of the events recorded by the detector within the energy range of $[300, 2700]\, \mbox{keV}$ are retained. The determination of this threshold value was conducted with the aim of optimizing the balance between achieving a satisfactory acceptance rate (calculated over the $[300, 2700]\, \mbox{keV}$ interval) and attaining a high ${}^{208}$Tl peak-to-Compton ratio, as depicted in Fig. \ref{fig:threshold_set}. We note that increasing the threshold value would further enhance the peak-to-Compton ratio, but this would come at the expense of a rapidly decreasing acceptance rate. The peak-to-Compton ratio has been calculated by taking the fraction of the events collected in the ${}^{208}$Tl full-peak region over the ${}^{208}$Tl single-Compton region, considering the energy intervals indicated in Table \ref{tab:training_regions}.

\begin{figure}
    \centering
    \includegraphics[width=0.95\linewidth]{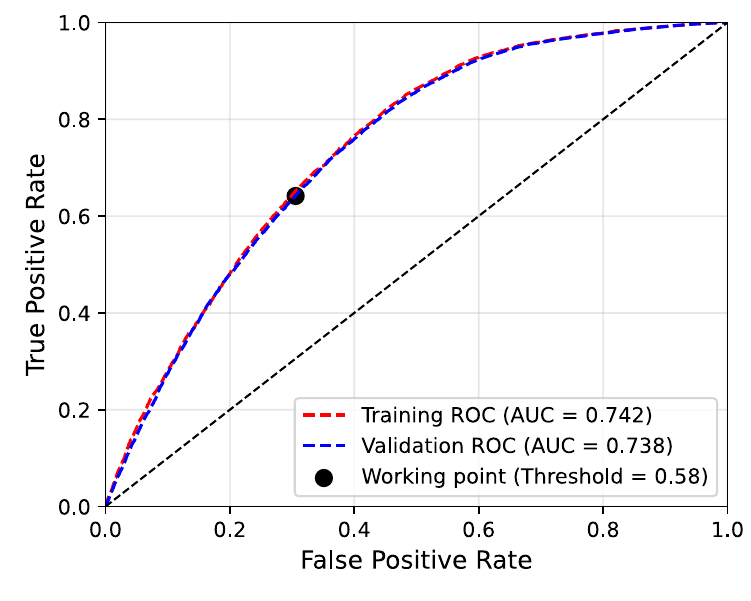}
    \caption{ROC curve for the Classifier on the labelled training and validation dataset partitions. Area Under Curve (AUC) and the TPR/FPR tradeoff at the working point are also shown.}
    \label{fig:classifier_roc}
\end{figure}

In Fig. \ref{fig:classifier_roc}, we present the Receiver Operating Characteristic (ROC) curve of the Classifier, illustrating the tradeoff between the True Positive Rate (TPR) and False Positive Rate (FPR). This curve is derived exclusively from data within labelled energy intervals. At the chosen working point, the Classifier retains approximately $50\%$ of the dataset, indicating how the network selectivity decreases (retains more events) inside the training energy regions.

In Sec. \ref{sec:cryoconite} we will show how the ACM model generalizes for a different dataset, outside the energy intervals used for training the Classifier, and we will compare the results with a novel, fully-unsupervised classification strategy described in Sec. \ref{sec:unsupervised}.

\section{Gaussian Mixture Model}\label{sec:gmm}
\label{sec:unsupervised}
In this section, we describe the Gaussian Mixture Model (GMM) for SSE-MSE classification \cite{GMM}. GMM is a statistical method that falls under the wider range of unsupervised techniques whose strength lies in not requiring data labeling during training. This technique is widely used across fields, including applications in high-energy physics \cite{WEBER2022166299} and astrophysics \cite{Weber_2022}. Using GMM to discriminate MSE from SSE might lead to an advantage with respect to the supervised method described in Sec. \ref{ACM_calibration}, as it relieves the need to label data. Data labeling as performed in the ACM method is, in fact, not exact, and restricted to narrow energy ranges. Furthermore, GMM does not require hyperparameter tuning, except for the number of clusters to be identified within the data.

\subsection{Model description}
A Gaussian mixture model is a fully unsupervised statistical method for identifying subpopulations inside an overall population assuming that the samples are normally distributed. Let $\boldsymbol{X}$ be a $n \times m$ data matrix where each row $\boldsymbol{x_i}$ is an observation of $m$ random variables. GMM models the observed probability density function as:
\begin{equation}
    p(\boldsymbol{x_i};\boldsymbol{\Theta})=\sum_{k=1}^{K}{\tau_k\phi(\boldsymbol{x_i};\boldsymbol{\mu_k},\boldsymbol{\Sigma_k}})
    \label{eq:GMMpdf}
\end{equation}
where $K$ is the number of gaussians employed, i.e.\ the number of clusters to be identified, $\tau_k$ are mixing weights and $\phi(\boldsymbol{x_i};\boldsymbol{\mu_k},\boldsymbol{\Sigma_k})$ is the pdf of a $m$ dimensional gaussian, with means $\mu_k$ and covariance matrices $\Sigma_k$:
\begin{equation}
    \begin{gathered}
        \phi(\boldsymbol{x_i};\boldsymbol{\mu_k},\boldsymbol{\Sigma_k}) = \\
        \frac{1}{\sqrt{(2\pi)^{m}\lvert\boldsymbol{\Sigma_k}\rvert}}\exp(-\frac{1}{2}(\boldsymbol{x_i}-\boldsymbol{\mu_k})^T\boldsymbol{\Sigma_k}^{-1}(\boldsymbol{x_i}-\boldsymbol{\mu_k}))
    \end{gathered} 
\label{eq:gaussian}
\end{equation}
The model can automatically estimate the mixing weights $\boldsymbol{\tau}$ and the parameters $\boldsymbol{\Theta}$ (corresponding to $\mu_k$ and $\Sigma_k$ in Eq. \ref{eq:GMMpdf} and Eq. \ref{eq:gaussian}) using the Expectation Maximization algorithm  \cite{Do2008} achieving a statistical description of the dataset.

\subsection{Dataset preprocessing}
\label{sub:GMM_preprocess}
For the unsupervised ML analysis, we adopted a traditional approach as described in previous studies \cite{Agostini2013, gerdaph2, Caldwell2015}, where features are engineered directly from the pulse shape using voltage and induced current data. This approach benefits from the interpretability of features grounded in physics intuition, which allows for a clear understanding of their role in distinguishing event classes. While using features extracted by an Autoencoder could potentially reduce information loss, we observed that the resulting class distributions did not form well-separated clusters (see Fig. \ref{fig:feature_distribution}), which limited their effectiveness for this purpose. More advanced feature extraction methods, such as Variational Autoencoders (VAEs) \cite{vae}, disentangled VAEs \cite{higgins2017betavae}, or the Uniform Manifold Approximation and Projection (UMAP) \cite{umap} technique, could potentially improve clustering and separability. Although these approaches lie beyond the scope of this study, they represent a promising direction for future exploration in this domain. We further comment on the use of these techniques in Sec. \ref{sec:conclusions}.

\begin{figure}[h!]
    \centering
    \includegraphics[width = 0.45\textwidth]{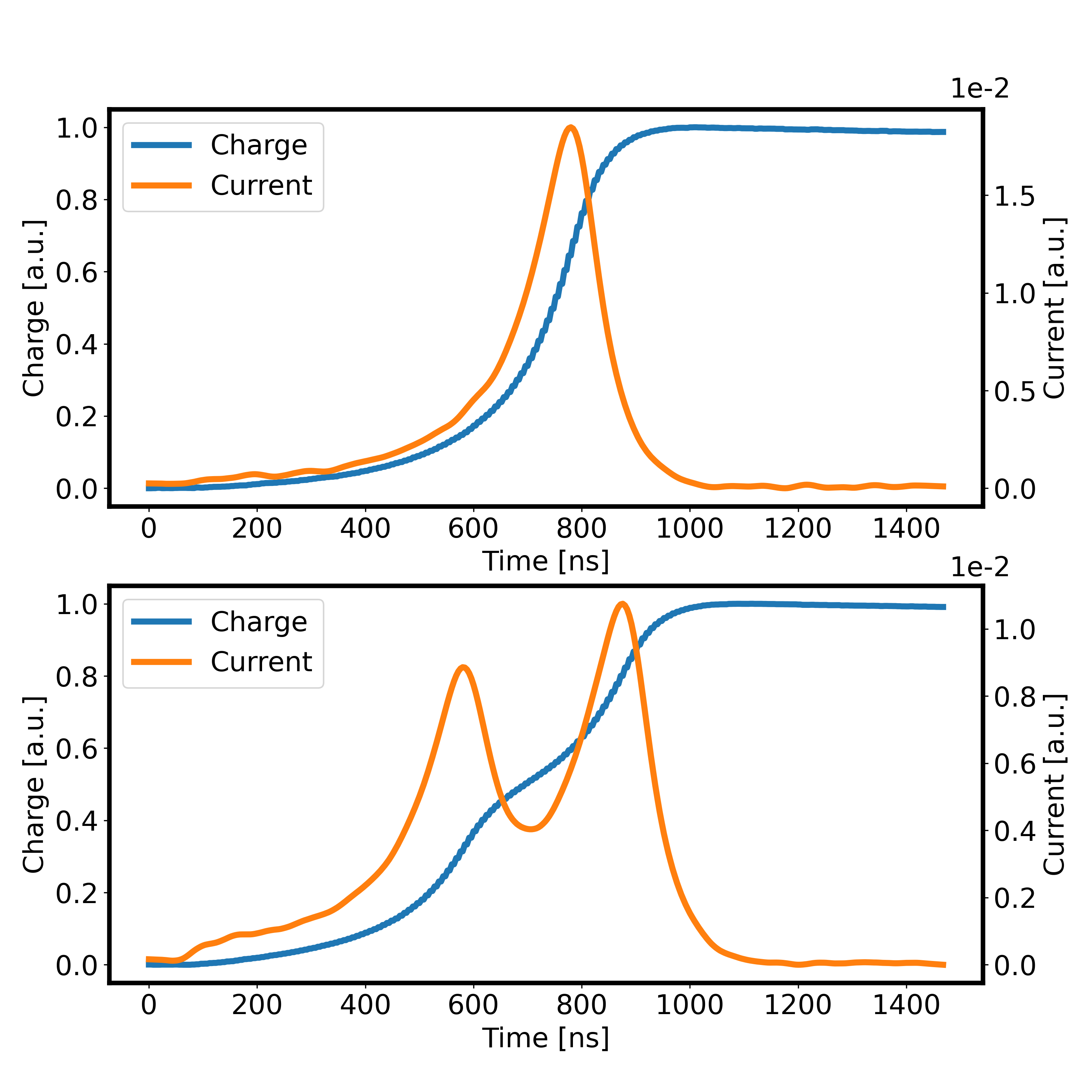}
    \caption{Action of two consecutive Savitzky-Golay filters on a typical SSE (top) and on a typical MSE (bottom). The two filters smooth ($19$ as window length and $2$ as polynomial order) and differentiate ($19$ as window length and $13$ as polynomial order) the signal respectively.}
    \label{fig:Experimental_events}
\end{figure}

The GMM features were by differentiating the voltage signal and applying a Savitzky-Golay smoothing filter~\cite{savitzky:filt}. Fig. \ref{fig:Experimental_events} shows the result of applying Savitzky-Golay filters to an example SSE and MSE.

\begin{figure*}
    \centering
    \includegraphics[width = \textwidth]{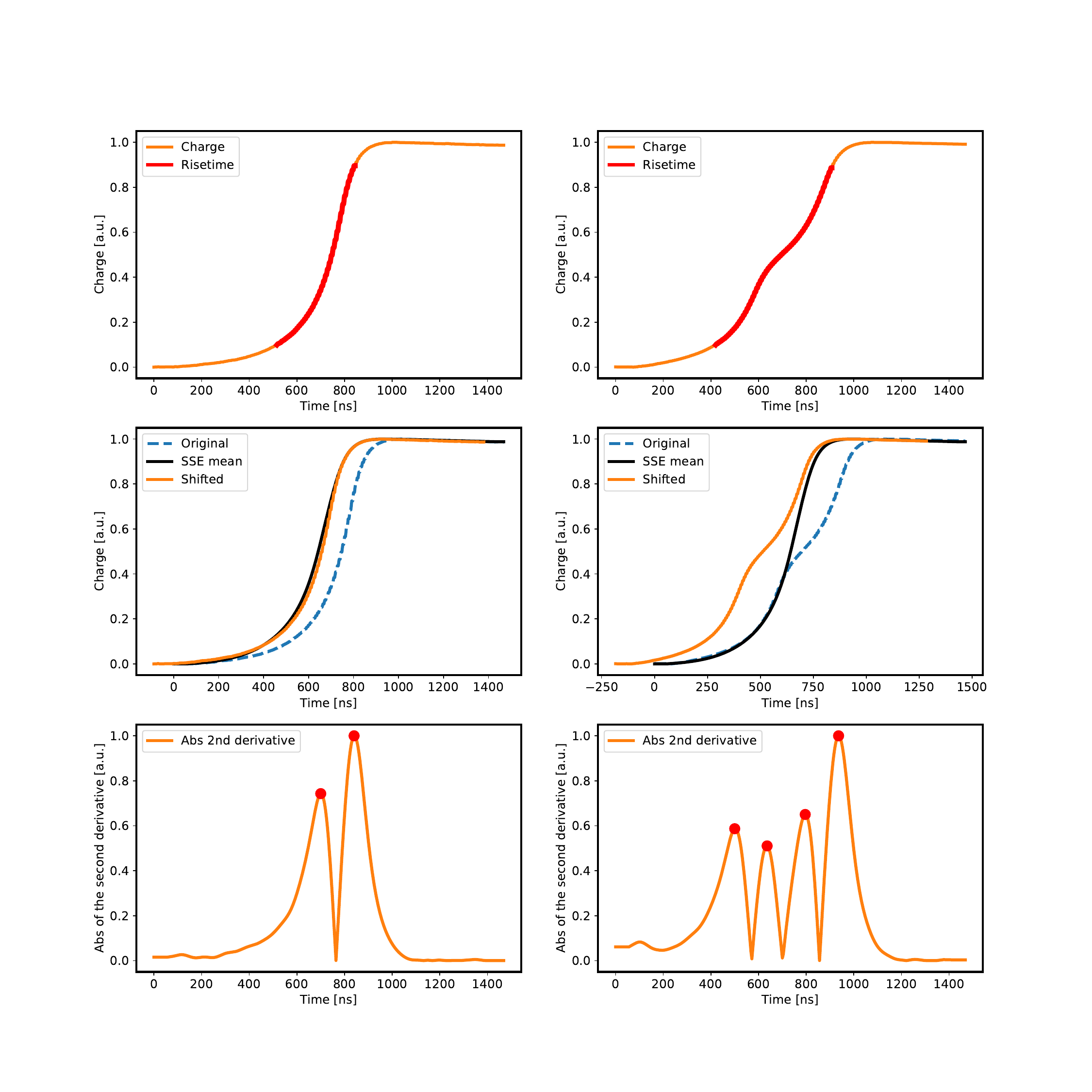}
    \caption{Representation of the difference between a typical SSE (first column) and a typical MSE (second column) concerning rise time (first row), L1 distance (second row) and number of peaks in the absolute value of the charge second derivative (third row). A more detailed explanation can be found in Sec. \ref{sub:GMM_preprocess}.}
    \label{fig:Comparison_of_events}
\end{figure*}

The unsupervised model makes use of three al\-go\-rit\-hmi\-cal\-ly-derived parameters:
\begin{itemize}
    \item Rise time of the pulse, defined as the time taken by the signal to go from its 10\% to 90\%. The mean rise time is expected to be higher for MSEs, as photon interactions from scattered gamma rays may occur in regions with varying weighting potential gradients. Consequently, the electron-hole pairs generated from the interaction have different collection times.
    \item L1 norm (or Manhattan distance) between each pulse and the average SSE pulse, that we obtained by averaging many pulses in regions where mostly SSEs are expected. The mean value of this parameter is expected to be higher for MSEs since their shape is influenced by charge collection. This effect becomes more evident as the energy increases.
    \item Number of peaks in the absolute value of the charge second derivative. The first derivative of the current is expected to have a maximum and a minimum for each energy deposit inside the crystal's active region. In principle, one could simply count the number of peaks in the current signal. However, this approach could lead to the misclassification of certain MSEs characterized by closely spaced energy deposits. In such cases, their current signal would not display two (or more) distinct peaks but rather a single peak preceded by a plateau. Counting the number of peaks in the absolute value of the second derivative of the charge instead enables better event identification.
\end{itemize}
Fig. \ref{fig:Comparison_of_events} gives an idea of the rationale behind the extracted parameters: the left column represents the three parameters for a typical SSE, while the right one for a typical MSE. On the top row, we compare the signal rise time which is expected to be longer for the MSE. In the second row, it can be observed that the L1 distance between the mean of several SSE events and the properly shifted SSE or MSE under consideration is expected to be lower for the SSE. In the bottom row, we compare the expected number of peaks of the second derivative of the charge absolute value for the two event classes. 

Before their extraction, input voltages were normalized from 0 to 1. A standardization was further applied to the extracted parameters to have 0 mean and unitary standard deviation. The obtained dataset consists thus of 1.5 million preamplified signals having energy in the range $[300, 2700]$ keV described by the three algorithmic derived features.

\subsection{Model validation}
\label{sub:GMM_validation}
\begin{figure}[h!]
    \centering
    \includegraphics[width = 0.45\textwidth]{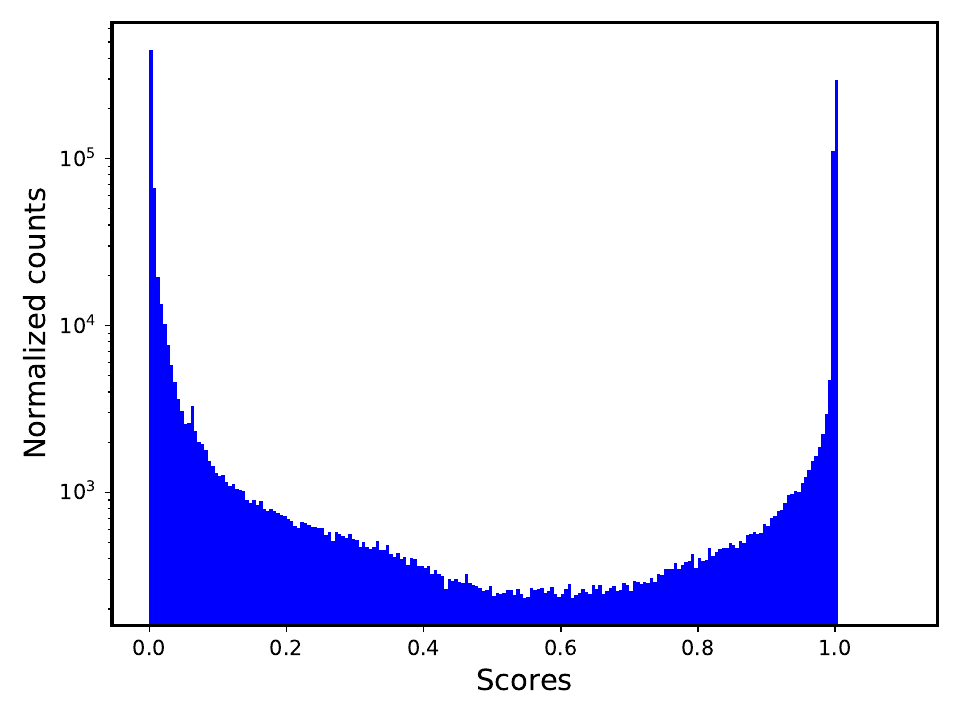}
    \caption{Distribution of the estimated scores for the GMM with $^{232}$Th source. The polarized distribution shows that a separation into two categories was achieved.}
    \label{fig:GMM_labels}
\end{figure}

The model used in this work was implemented using the open-source library \texttt{Scikit-learn} \cite{scikit-learn}, version 1.2.2. To properly distinguish MSE from SSE the number of Gaussians $K$ was set to two, specifying the covariance to be "full" to leave to the algorithm the maximum number of degrees of freedom on the choice of the Gaussian shapes. This choice is justified because hand-extracted features may in principle be codependent\footnote{The model takes on average $8.45$ seconds to fit using Intel\textsuperscript{\tiny\textregistered}Xeon\textsuperscript{\tiny\textregistered} CPU (2.20 GHz).}. After fitting the model, the component densities for each sample were available. Since we are dealing with a binary classification problem, the probability that each sample belongs to one of the Gaussians is completely determined by the other, allowing us to use one of them as a prediction score. Fig. \ref{fig:GMM_labels} shows the distribution of the labels obtained for the $^{232}$Th source: as one can see, a polarization of the data set was achieved as desired.

The unsupervised nature of the problem doesn't allow us to establish which score (0 or 1) corresponds to the MSE class we are interested in. Thresholding the distribution at 0.5 allowed us to solve that indetermination by checking for which events a reduction of the SSE regions was achieved. Concerning the score distributions represented in Fig. \ref{fig:GMM_labels}, every event with a label not greater than 0.5 was marked as MSE. Since no label was involved, the classifier characterization in terms of a ROC curve (similarly to the Classifier characterization in section \ref{ACM_calibration}) could not be carried out. The silhouette score \cite{Silhouette} has been evaluated, which is a measure of cohesion and separation between the elements of the two clusters. Evaluating such a score requires the distance matrix, achieved by calculating the distance between every point from each other. Since this operation is time expensive, the silhouette score has been evaluated only considering $10^4$ events (0.09\% of the total) randomly selected while maintaining the proportion of events belonging to the two clusters. The achieved silhouette score is $0.29$, which may indicate that the clusters are not completely separated for the events observed with the $^{232}$Th source.

\section{Comparison on a Cryoconite dataset}
\label{sec:cryoconite}

\begin{figure*}[!h]
    \centering
    \includegraphics[width = \textwidth]{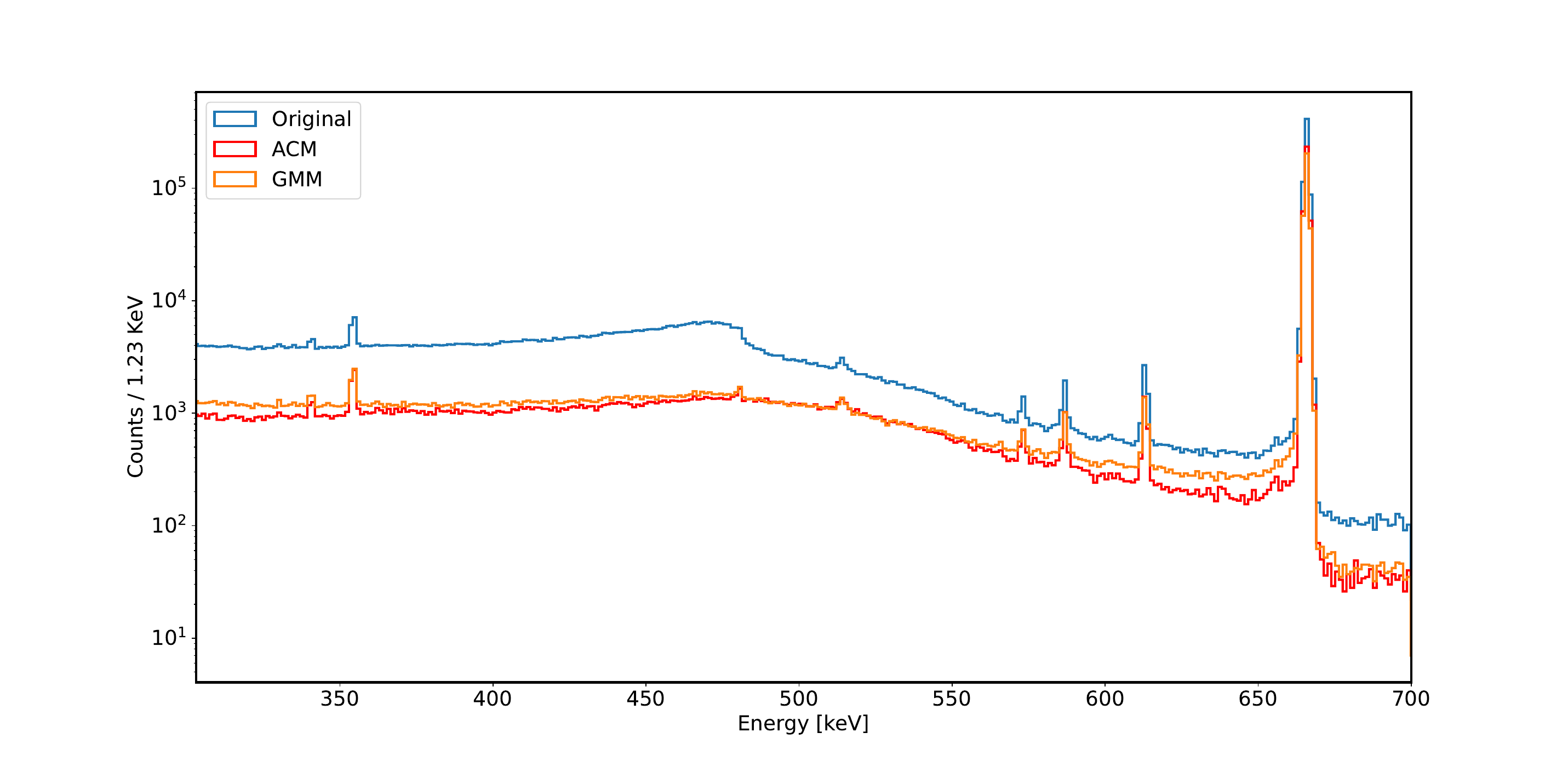}
    \caption{Comparison between the original energy spectrum (blue), the one cut through ACM (red) and the one cut through GMM (orange) for the considered cryoconite sample.}
    \label{fig:spectrum}
\end{figure*}

In this section, the supervised (ACM) and unsupervised (GMM) techniques are compared on an environmental sample that was never used during the training phase of the ACM. For this sample, we kept the classifier score of the ACM at 0.58, while for the GMM we fixed the threshold at 0.5\footnote{The silhouette score achieved with with this dataset is $0.43$.}. Fig. \ref{fig:spectrum} shows the original energy spectrum and the result of ACM and GMM selections in the energy range between $300$ and $700$ keV.

\begin{figure}[!h]
    \centering
    \includegraphics[width = 0.45\textwidth]{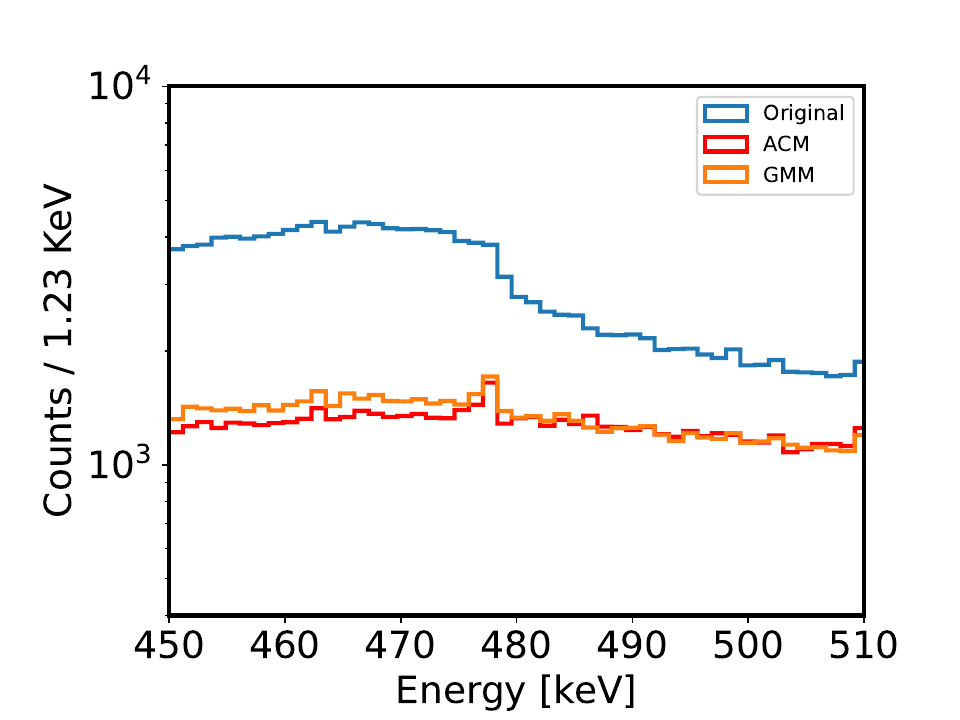}
    \caption{Close-up view between 450 keV and 510 keV. Around 477 keV the peak of $^{7}$Be emerges for both ACM and GMM.}
    \label{fig:inlet}
\end{figure}

We observe that both techniques were able to suppress the Compton background while retaining approximately 56.6\% (ACM) and 49.5\% (GMM) of the events under the $^{137}$Cs full-energy peak region. More significantly, in the Compton continuum region, the discrimination resulted in discarding 75\% of counts for ACM and 72\% for GMM, allowing the underlying peaks to emerge. Remarkably, in the region around $477\,\mbox{keV}$, a barely visible peak in the original spectrum stands out, which we attribute to the ${}^7$Be full-energy peak. The identification of ${}^7$Be agrees with previous results showing that this natural short-lived ($t_{1/2}=53.3$ d) fallout radionuclide is found in excess in cryoconite because of exchanges occurred between cryoconite and glacial meltwater in the weeks before sampling \cite{Beard2024}. ${}^7$Be is typically detectable if counting is carried out a few months after sampling, before the nuclide has completely decayed. The method for background suppression proposed here allowed us to partially extend this interval, as between sampling and counting $30$ months passed.
Fig. \ref{fig:inlet} provides an inset of this region. To quantitatively assess the effectiveness of the Compton-background reduction, the ratio between the $^{137}$Cs full-energy peak counts and the Compton background in the energy range from $355\,\mbox{keV}$ to $450\,\mbox{keV}$ was calculated and is presented in Table \ref{tab:peak_compton}.

\begin{table}[ht]
    \centering
    \caption{Table containing ratios between counts under $^{137}$Cs full-energy peak and Compton background in the $[355, \, 450]$ keV energy range.}
  \begin{tabular}{ccc}
    \toprule
    Original spectrum & ACM cut & GMM cut \\
    \midrule
    0.238 & 0.547 & 0.414 \\
    \bottomrule
  \end{tabular}
  \label{tab:peak_compton}
\end{table}
The peak count estimates required to compute the values in table \ref{tab:peak_compton} were obtained by fitting the $^{137}$Cs peaks in the spectra in Fig. \ref{fig:spectrum} with a Gaussian function plus a step function to account for the abrupt change in the background. To estimate the Compton background we summed the events in the $[355, \, 450]$ keV energy interval. Both ACM and GMM post-selected histograms achieved an enhanced signal-to-background ratio.

\section{Conclusions}
\label{sec:conclusions}
BEGe\texttrademark{} detectors are powerful devices in gamma-ray spectroscopy, benefitting from excellent energy resolution and allowing us to apply powerful pulse-shape discrimination techniques that mitigate background events originating from Compton scattering inside the crystal. In this study, we presented two machine learning methods that achieved effective single-site and multi-site event classification through fully data-driven approaches, without resorting to simulations. Such analysis tools are cost-effective and are easily implementable on any commercial detector.

The first method consisted of two feed-forward neural networks, an autoencoder for data compression and a classifier for event prediction. Notably, the classifier was able to generalize its predictions for events outside the initially narrow, physics-informed training regions. The second model employed an unsupervised Gaussian Mixture Model approach, successfully implementing Single-Site Event (SSE) rejection through clustering elementary functions of the detector signal waveform. This demonstrates the effectiveness of a new Machine Learning model in the context of SSE rejection in gamma-ray spectroscopy, which is fully unsupervised and requires simple and almost physics-agnostic data preprocessing.
Both models succeeded in identifying a peak at $477\,\mbox{keV}$, which was submerged by the Compton background of the $^{137}$Cs peak in the spectrum before selection. Both models revealed the $477\,\mbox{keV}$ peak with comparable performance, the supervised method being slightly more efficient in enhancing the $^{137}$Cs peak-to-Compton ratio.

While this study employed a standard Autoencoder, future work could investigate the use of a VAE or a disentangled VAE to generate more expressive and interpretable latent embeddings, as suggested in Sec. \ref{sec:acm}. By introducing a probabilistic framework, VAEs impose a structured prior on the latent space, potentially improving generalizability and robustness to input variations. Disentangled VAEs, in particular, can offer even greater interpretability by learning independent latent features, which may help isolate the characteristics most relevant to MSE/SSE discrimination. This could enhance the model performance, especially when generalizing to datasets with varying energy distributions. Furthermore, training clustering models on VAE-extracted features rather than manually engineered ones (which are more prone to information loss) might further enhance the unsupervised strategy, as more information could be retained from the original waveform and Contrastive Learning could be implemented \cite{bai2022gaussianmixturevariationalautoencoder}. Other possible directions involve leveraging UMAP to achieve the desired dimensionality reduction, and exploring non-para\-met\-ric clustering methods such as Density Peaks Advanced \cite{DERRICO2021476} to relieve the requirement of prior knowledge on the number of clusters.

This study bears significance in evaluating the effectiveness of machine learning strategies in gamma-ray spectroscopy, with implications extending to various applications such as nuclear physics, Fundamental physics, and medical imaging. The findings acknowledge and motivate further exploration towards machine learning strategies to enhance data analysis for BEGe\texttrademark{} gamma-ray spectroscopy.
\\

{\footnotesize \noindent
\textbf{Data availability statement} The da\-ta\-set produced and analysed during this study and the codebase are available upon reasonable request from the authors.
}

\begin{acknowledgements}
AG acknowledges support by the Horizon 2020 Marie Sk\l{}odowska-Curie actions (H2020-MSCA-IF GA No.101027746). We acknowledge the INFN electronics workshop team for their support in providing the warm amplifier. We would like to extend our thanks to Giovanni Benato and Claudio Gotti for their valuable contributions and insightful discussions. Mention of commercial products is for information only and does not imply recommendation or endorsement.
\end{acknowledgements}

\bibliographystyle{spphys}       
\bibliography{begeml}   

\end{document}